\documentclass[twocolumn,prd,groupedaddress,nofootinbib,showpacs]
{revtex4}
\usepackage{latexsym}
\usepackage{amssymb}
\usepackage{graphicx}
\usepackage{dcolumn}
\usepackage{bm}

\newcommand{\ii}{\'{\i}}

\newcommand{\zero}{\setcounter{equation}{0}}

\begin{document}

\title{The Noncommutative  U(N) Kalb-Ramond  Theory}

\author{R. Amorim$^{1, a}$, C.N. Ferreira$^{1, 2, b}$ \footnote{crisnfer@cbpf.br}
and C.F.L. Godinho$^{2, 3, c}$}

\affiliation{$^1$Instituto de F\'{\i}sica, Universidade Federal do
Rio de Janeiro
POBOX 68528, 21945-910, Rio de Janeiro, RJ, Brazil\\
$^2$Grupo de F\'{\i}sica
Te\'orica Jos\'e Leite Lopes, Petr\'opolis, RJ, Brazil\\
$^3$ Centro Brasileiro de Pesquisas F\'{\i}sicas (CBPF-CCP), Rua
Dr. Xavier Sigaud 150, Urca 22.290-180, Rio de Janeiro, RJ,
Brazil}

\begin{abstract}
We present the noncommutative extention of the U(N)
Cremmer-\-Scherk-\-Kalb-\-Ramond  theory, displaying
its differential form and gauge structures.
The Seiberg-Witten map of the
model is also constructed up to $0(\theta^2)$.
\end{abstract}

\maketitle

\section{Introduction}

The first ideas about  space-time noncommutativity were formulated by Heisenberg in the thirties \cite{Heis},
although the first published work on the subject appeared in 1947 \cite{Sn}, introducing a possible framework for avoiding the  characteristic singularities  
of quantum field theories.  Although the original motivations
have been eclipsed by
the renormalization program,
recently the interest on noncommutative theories has
grown up in large scale, mainly associating    the   space-time
noncommutativity with results coming from string theory . In this
approach noncommutativity can
be found in the study of perturbative strings
in the presence of the branes  in a constant
background magnetic field \cite{PO,HV,CH1}. For a review on the subject, 
where several interesting features of noncommutative field theories can be commented,
see for instance \cite{REVIEW} and references therein. 
One of the  important ingredients of noncommutative gauge field theories 
is the Seiberg-Witten map connecting field variables which transform under a noncommutative gauge structure with
ordinary field variables transforming under an ordinary gauge structure
\cite{Seiberg,AK,AF,ABF}. This map seems to be essential for the construction of
a phenomenological viable noncommutative description of Nature \cite{WESS,CALMET}.
\medskip

Interesting models of gauge theories that have not had their noncommutative extentions very
explored are those constructed with the aid of antisymmetric tensor fields.
The antisymmetric  Kalb-Ramond tensor field  has been first introduced within a string  theory\cite{KR,CS,N76},
but the so called  Cremmer-Scherk-Kalb-Ramond  (CSKR) model appears in a great variety of scenarios,
including supersymmetric theories\cite{Gates,Vilenkin,Baulieu}, cosmology\cite{Gasperini} and 
cosmic strings\cite{CHP2002}. 
Other interesting points are  related to the rank of the CSKR model gauge structure and its consequences 
under quantization, including the possibility of mass for gauge theories without spoiling gauge invariance
\cite{Henneaux,Lahiri,AB}. 

\medskip

In the present work, we propose a noncommutative  generalization of the 
CSKR theory . We show that its covariant description, with
the aid of differential forms, can be extended in order to incorporate  Moyal products,
characteristic of noncommutative field theories.
It is also possible to show that there exits an underlying
commutative gauge invariant theory and a suitable Seiberg-Witten map
linking the noncommutative model and its ordinary counterpart.

\medskip
The outline
of this paper is as follows: in Section 2, we start by presenting
the ordinary U(N) CSKR model in terms of differential forms. After that
we show that it is possible to deform the form structure in order to incorporate 
Moyal products.
This essentially permits the construction of the noncommutative extention of the model.
In Section 3, the appropriate Seiberg-Witten map is derived. It takes into account not only the usual Yang-Mills sector but also
the gauge sector which arises when one considers the  invariance associated with the 1-form gauge parameters.
We reserve
Section 4 for some concluding remarks.

\section{The noncommutative U(N)  CSKR model \zero }

\noindent

\noindent

To fix notations and conventions,
let us start  with a brief review of the non Abelian
commutative Kalb-Ramond theory.  We will follow a notation 
close to the one found in Ref. \cite{AB}. After that we   will study  the corresponding
noncommutative theory. 

Let
 $a = a_{\mu}^a T^a dx^{\mu }$
represent a one-form connection taking values in the $U(N)$ algebra  in the fundamental 
representation. We assume that

\begin{eqnarray}
[T^a, T^b] &=& i f^{abc} T^c\nonumber\\
\{ T^a, T^b\} &=& d^{a b c} T^c \nonumber\\
 tr (T^a T^b) &=& \frac{1}{2} \delta^{ab}
\label{2.2}
\end{eqnarray}

On any $u(N)$ valued p-form $\alpha $
it is possible to define the exterior covariant derivative 

\begin{equation}
D\alpha = d \alpha - i a \wedge \alpha + i (-1)^p \alpha \wedge a
\label{2.4}
\end{equation}

It follows the first Bianchi identity

\begin{equation}
DD\alpha=i[\alpha,f]
\label{2.5}
\end{equation}

\noindent where we have defined  the curvature two-form 

\begin{equation}
f = da -i a \wedge a
\label{2.6}
\end{equation}

\noindent and the wedge product is implicit in the commutator.
For completeness, we also note the second Bianchi identity

\begin{equation}
Df=0
\label{2.7 }
\end{equation}

The gauge sector of the $U(N)$ Yang-Mills theory, with this notation, is described by the action

\begin{equation}
S=Tr \int  f \,\, \; \; ^{\diamond}\! \! \! \! \! \! \!\! \wedge  f
\label{2.8}
\end{equation}

\noindent where the symbol $\diamond$ here denotes the space-time dual. Action (\ref{2.8}) is invariant under
the gauge transformation

\begin{equation}
\bar\delta a= D\alpha
\label{2.9}
\end{equation}

\noindent since under (\ref{2.9}) $f$ transforms as

\begin{equation}
\bar\delta f= i[\alpha,f]
\label{2.10}
\end{equation}

\noindent and the invariance of (\ref{2.8}) is achieved  due to the cyclic property of the trace operation.
In the above expressions we have used $\bar\delta$  to represent ordinary gauge variations. We will let $\delta$ represent the corresponding noncommutative gauge variations .
\medskip

To describe the  CSKR model, besides the connection $a$, we need a two-form gauge field
$b=\frac{1}{2} b_{\mu \nu} dx^{\mu}\wedge dx^{\nu}$
and a compensating one-form field
$\omega=\omega_\mu dx^\mu$, both taking values in $u(N)$.
The 3-form field strength associated with $b$ is defined as

\begin{equation}
g=Db
\label{2.12}
\end{equation}

\noindent Now the  CSKR action

\begin{equation}
S=Tr \int \left[   f \,\, \; \; ^{\diamond}\! \! \! \! \! \! \!\! \wedge  f - 
 \hat g \,\, \; \; ^{\diamond}\! \! \! \! \! \! \!\! \wedge
\hat g  + 2 m f \wedge \hat b \right] \label{2.13}
\end{equation}

\noindent shows itself to be invariant under the set of gauge transformations

\begin{equation}
\begin{array}{ll}
\bar\delta {a} = D\alpha\\
\bar\delta  b  = D \xi  + i[\alpha , b]\\ 
\bar\delta \omega  = i[\alpha , \omega ]   - \xi \
\end{array}
\label{2.14}
\end{equation}

\noindent where we have used in (\ref{2.13}) the collective two-form

\begin{equation}
\hat b = b + D \omega
\label{2.15}
\end{equation}

\noindent  and the modified field strength

\begin{equation}
\hat g = D \hat b 
\label{2.16}
\end{equation}

We observe from (\ref{2.14}) and the definitions above that
$b$ and $\omega$ transform not only as Yang-Mills tensors but also present an additional
transformation related to the one-form gauge parameter $\xi=\xi_\mu dx^\mu$. The quantities $\hat b$ and $\hat g$, however, transform only as
Yang-Mills tensors, in the same way as $f$ in (\ref{2.10}). This fact permits the gauge invariance of (\ref{2.13}). Actually, in the Abelian case,
the compensating one-form $\omega$ is not necessary, and the corresponding theory is gauge invariant,
although reducible\cite{Henneaux,Lahiri,AB}. 

All of the transformations defined above close in an algebra, defined by the parameters composition rule

\begin{eqnarray}
\alpha_3&=&i[\alpha_2,\alpha_1]\nonumber\\
\xi_3&=& i[\xi_2,\alpha_1]-i[\xi_1,\alpha_2]
\label{2.17}
\end{eqnarray}

\noindent when the commutation of two successive gauge transformations is applied to any one of the fields appearing in the theory,
here generically represented by $y$:

\begin{equation}
[\bar\delta_1,\bar\delta_2] y=\bar\delta_3 y
\label{2.18}
\end{equation}

The field-antifield quantization of the model described above was studied in Ref. \cite{AB}.
\medskip

Let us now pass to consider the noncommutative version of this theory. 
As already commented, the basic procedure to construct the noncommutative
extension of some  theory consists in deforming ordinary products
to noncommutative Moyal products.
For any two fields $\Phi_1(x)$ and $ \Phi_2(x)$, we define
their Moyal product as

\begin{equation}
\Phi_1(x) \star  \Phi_2(x) = exp\left(\frac{i}{2} \theta^{\mu \nu}
\partial_{\mu}^{x} \partial^y_{\nu}\right) \Phi_1(x)\Phi_2(y)|_{x=y}
\label{2.19}
\end{equation}

\noindent
where $\theta^{\mu \nu}$ is assumed to be a real, constant and antisymmetric quantity
which characterizes the noncommutativity of the theory. These products are associative and cyclic under the integral sign,
if adequate boundary conditions are assumed.
As it is well known \cite{Seiberg,AF},  the $U(N)$ group elements  are also deformed by such a product
in the sense that their construction  by exponentiation involves Moyal products. Also the group multiplication is defined as a Moyal
product. 
In this way the symmetry 
structure of the noncommutative $U(N)$ theory is not the same as the corresponding commutative one and the group
closure property is only achieved if the algebra generators close not only under commutations but also under anticommutation. This essentially constitutes the reason for choosing $U(N)$ in place of $SU(N)$ as a symmetry group
of this noncommutative gauge theory, although other possibilities can be considered \cite{WESS}.
Similar deformations can also be implemented in the differential forms structure.
In this way, the exterior product is modified in order to accommodate the Moyal structure with the  formal replacement
 $   \wedge \rightarrow \,\,^{^{^{\star}}}\! \! \! \!   \wedge \,\, $.
In a coordinate basis, this modification  is trivial and consists in introducing Moyal star products in place of the ordinary ones between the forms components, keeping the wedge product between the form basis. Here we will restrict ourselves to this situation.
Similar procedures can  also be implemented in the definition of the exterior covariant derivative. 
The  non-commutative version
of the exterior covariant derivative of a p-form $\Lambda$ is given by

\begin{equation}
D\Lambda = d \Lambda - i A \; \; ^{^{^{\star}}}\! \! \! \! \!  \wedge \Lambda +
 i (-1)^p \Lambda \; \; ^{^{^{\star}}}\! \! \! \! \!   \wedge A
\label{2.20}
\end{equation}

\noindent Definition (\ref{2.6}) is in the same way trivially deformed to

\begin{eqnarray}
F &=&dA -i A  \; \; ^{^{^{\star}}}\! \! \! \!\! \wedge A\nonumber\\
&=&\frac{1}{2}(( \partial_{\mu} A_{\nu}^a - \partial_{\nu} A_{\mu}^a) T^a 
-i [ A_{\mu}^a T^a \; \; ^{\star}\! \!, A_{\nu}^b T^b] )dx^{\mu} \wedge dx^{\nu}\nonumber\\
&=& \frac{1}{2}F_{\mu\nu}dx^{\mu} \wedge dx^{\nu}
\label{2.21}
\end{eqnarray}

\noindent where $A = A^a_{\mu}T^adx^{\mu}$ now represents the noncommutative 1-form connection.
The above expression  shows the rule played by the noncommutative wedge product  $\,\,\,^{^{^{\star}}}\! \! \! \!   \wedge \,$.
As can be observed, $F$ involves both structure functions defined in (\ref{2.2}).
Actually, $F_{\mu\nu}^a=\partial_{\mu} A_{\nu}^a - \partial_{\nu}A^a_\mu+ \frac{1}{2}f^{abc}\{A _{\mu}^b \; \; ^{\star}\! \!, A_{\nu}^c\} - \frac{i}{2}d^{abc}[A_{\mu}^b  \; \; ^{\star}\! \!, A_{\nu}^c]$. 
Other expressions follow the same rules. 
The Bianchi identities are now written as

\begin{eqnarray}
DD\Lambda&=&i[\Lambda\buildrel\star\over,F]\nonumber\\
DF&=&0
\label{2.22}
\end{eqnarray}

To construct the noncommutative version of the model described above,
in place of the ordinary quantities $b$ and  $\omega$ we define the noncommutative forms 
$B= \frac{1}{2} B_{\mu \nu} dx^{\mu}\wedge dx^{\nu}$ and
$\Omega= \Omega_\mu dx^\mu$.
Also in place of
$g$, $\hat b$ and $\hat g$ defined in 
(\ref{2.12}), (\ref{2.15}) and (\ref{2.16}), we introduce the corresponding noncommutative forms

\begin{eqnarray}
G&=&DB\nonumber\\
\hat B&=&B+D\Omega\nonumber\\
\hat G &=& D \hat B
\label{2.242}
\end{eqnarray}

\noindent These quantities present the following set of gauge transformations:

\begin{equation}
\begin{array}{ll}
\delta {A} = D\epsilon \\
\delta  B  = D \Xi  + i[\epsilon \; ^{\star}\! \!,  B] \\
\delta G= i[\Xi  \; \; ^{\star}\! \!, F] + i[ \epsilon  \; \; ^{\star}\! \!, G]  \\
\delta \Omega  = i[\epsilon \; ^{\star}\! \!,  \Omega ]   - \Xi \\
\delta F= i [\epsilon \; \; ^{\star}\! \!,  F  ]   \\
\delta {\hat B} = i[\epsilon \;\; ^{\star}\! \!,  {\hat B}]\\
\delta {\hat G}=  i [\epsilon  \; \; ^{\star}\! \!,  \hat G ]
\end{array}\label{2.241}
\end{equation}

\noindent and, as can be verified, the noncommutative extension of action (\ref{2.13}),

\begin{equation}
S=Tr \int \left[  F  \; \; ^{\diamond}\! \! \! \! \! \! \! \wedge  F -  \hat G \; \; ^{\diamond}\! \! \! \! \! \! \! \wedge
\hat G  + 2 m F \wedge \hat B \right] \label{2.25}
\end{equation}

\noindent is gauge invariant under (\ref{2.241}) due to cyclic properties of the Moyal product under the integral sign.
Due to the boundary conditions we are adopting, it is irrelevant to use
in the above expression $   \wedge $ or $\,\,^{^{^{\star}}}\! \! \! \!   \wedge \,\, $.

The transformations  (\ref{2.241}) close in an algebra

\begin{equation}
[\delta_1,\delta_2] Y=\delta_3 Y
\label{2.26}
\end{equation}

with the composition rule given by

\begin{eqnarray}
\epsilon_3&=&i[\epsilon_2\buildrel\star\over,\epsilon_1]\nonumber\\
\Xi_3&=& i[\Xi_2\buildrel\star\over,\epsilon_1]-i[\Xi_1\buildrel\star\over,\epsilon_2]
\label{2.27}
\end{eqnarray}

\noindent in place of (\ref{2.17}) and (\ref{2.18}).

As one can observe, the noncommutative extention of the CSKR theory has been constructed
without difficulties. Of course its quantum version would show all those characteristic points
associated with noncommutative field theories and their non planar diagramatic expansions \cite{REVIEW}.
We will not consider these points in this work.
In what follows let us study the Seiberg-Witten map of the model, which presents some interesting features.

\section{The Seiberg-Witten map \zero}

Accordingly to what we have been discussing in the last section,
let us represent the 
noncommutative field variables by capital letters, here
generically denoted by $Y$ and the corresponding ordinary ones by 
small letters generically  written as
$y$. Also accordingly to the notations of the last section,
their gauge transformations are  respectively represented  by $\delta Y$ and $\bar\delta y$.
The basic idea in
the construction of the Seiberg-Witten map is to obtain the gauge transformations $\delta Y$ of the noncommutative 
variables  departing from the gauge 
structure of the  ordinary theory, with variables transforming accordingly to $\bar\delta y$. This is equivalent to solve the equation

\begin{equation}
\delta Y=\bar\delta Y[y]
\label{SW1}
\end{equation}

\noindent which is done by using expansions in powers of the noncommutative parameter $\theta$
and assuming that up to $0(\theta^2)$ terms, $Y\rightarrow y$.
This map is non trivial when  the noncommutative parameters $\epsilon$ and $\Xi$ are considered as functions
of the commutative parameters $\alpha$ and $\xi$ as well as of the ordinary fields $y$.
In this case, the  fundamental expressions implicit in  (\ref{2.26}) reduce to

\begin{eqnarray}
\left[\,\bar\delta_1,\bar\delta_2\,\right]\,  A [y]
&=&D\epsilon_3[y]\nonumber\\
\left[\,\bar\delta_1,\bar\delta_2\,\right]\, \Omega [y]
&=& i[\epsilon_3[y]\buildrel\star\over,\Omega]-\Xi_3[y]    \nonumber\\
\left[\,\bar\delta_1,\bar\delta_2\,\right]\, B [y]&=&i[\epsilon_3[y]\buildrel\star\over,B]+D\Xi_3[y]
\label{SW2}
\end{eqnarray}

\noindent where now, in place of (\ref{2.27}), we get

\begin{eqnarray}
\label{SW3}
\epsilon_3[y]&=&
\bar\delta_1\epsilon_2[y]-
\bar\delta_2\epsilon_1[y]+
i\left[\epsilon_2[y]\buildrel\star\over,\epsilon_1[y]\right]\nonumber\\
\Xi_3[y]&=& \bar \delta_1\Xi_2 - \bar \delta_2\Xi_1 
+ i[\epsilon_2 \; ^{\star}\! \!, \Xi_1] 
- i[\epsilon_1 \; ^{\star}\! \!, \Xi_2]\nonumber\\
\end{eqnarray}

\noindent due to the dependence of the parameters in the fields. 
Indices $1$, $2$ and $3$ represent the dependence of 
$\epsilon$ and $\Xi$ in $\alpha_i$ and $\xi_i$, $i=1,2,3$. For instance, 
$\epsilon_3[y]\equiv\epsilon[\alpha_3,\xi_3,y]$. Related quantities such  as $F$, $\hat B$ or $\hat G$
also follow similar rules related to the closure of the algebra.
\medskip

The first of equations (\ref{SW2}) is not new in the literature \cite{Seiberg,AK,WESS}, what does not occur with the second one, as far as we know.  They will be important
for the results that we will derive.
As in the pure Yang-Mills case,  the gauge transformation 0-form parameter $\epsilon$
can be expanded to first order in $\theta^{\mu\nu}$ as
$\epsilon[y]=\alpha+\epsilon^{(1)}[y]$. In the same way, the 1-form parameter $\Xi$ is expanded 
as $\Xi[y]=\xi+\Xi^{(1)}[y]$, to first order in $\theta$. 
From (\ref{SW3}) and the above expansions we deduce that

\begin{eqnarray}
\bar\delta_1\epsilon^{(1)}_2-\bar\delta_2\epsilon^{(1)}_1
&-&i[\alpha_1,\epsilon^{(1)}_2]+i[\alpha_2,\epsilon^{(1)}_1]
- \epsilon^{(1)}_3 \nonumber\\
&=&-\frac{1}{2}\theta^{\mu\nu}\{\partial_\mu\alpha_1,\partial_\nu\alpha_2\}
\label{SW4}
\end{eqnarray}

\noindent and also that

\begin{eqnarray}
 \bar\delta_1 \Xi_2^{(1)} &-&\bar \delta_2 \Xi_1^{(1)}  + i [\alpha_2, \Xi_1^{(1)}]  - i [\alpha_1, \Xi_2^{(1)}] -\Xi_3^{(1)}\nonumber\\
&=& -i [\epsilon_2^{(1)}, \xi_1] + i [\epsilon_1^{(1)}, \xi_2]\nonumber\\
&+&\frac{1}{2}\theta^{\alpha \beta }\Big(\{\partial_\alpha \alpha_2,\partial_\beta \xi_1\}
 - \{\partial_{\alpha} \alpha_1, \partial_{\beta} \xi_2\}\Big) 
\label{SW5}
\end{eqnarray}

The general solution of (\ref{SW4}) given by \cite{AK}

\begin{equation}
\epsilon^{(1)} =  \frac{\theta^{\alpha \beta}}{4} \{\partial_{\alpha}\alpha, a_{\beta}\} +
\lambda\theta^{\alpha \beta} [\partial_{\alpha}\alpha, a_{\beta}]
\label{SW6}
\end{equation}

\noindent where the term in $\lambda$ is the solution of the homogeneous part of (\ref{SW4}). Now it is possible to show that

\begin{equation}
\Xi^{(1)} = \frac{1}{2} \theta^{\alpha \beta}
\{ a_{\alpha}, \partial_{\beta} \xi + \frac{i}{2} [ a_{\beta}, \xi]\}
+i\lambda \theta^{\alpha \beta}[a_\alpha[a_\beta,\xi]]
\label{SW7}
\end{equation}

\noindent solves (\ref{SW5}) when one uses (\ref{SW6}) for $\epsilon^{(1)}$. 
Once we expand the fields to first order in $\theta$, this is: $A=a+A^{(1)}$, $\Omega=\omega+\Omega^{(1)}$ and $B=b+B^{(1)}$, we can rewrite the corresponding gauge transformations appearing in (\ref{2.241}) as

\begin{eqnarray}
\bar\delta A^{(1)}-i[\alpha,A^{(1)}]&=&-\frac{1}{2}\theta^{\alpha\beta}
\{\partial_\alpha\,\alpha,\partial_\beta a\}+D \epsilon^{(1)} \nonumber \\
\bar \delta \Omega^{(1)} - i [ \alpha , \Omega^{(1)}]&=& - \frac{1}{2} \theta^{\alpha \beta } \{ \partial_{\alpha} \alpha, \partial_{\beta} \omega \}  + i [\epsilon^{(1)}, \omega ] - \Xi^{(1)} \nonumber \\
\bar \delta B^{(1)} - i [ \alpha , B^{(1)}]&=& - \frac{1}{2} \theta^{\alpha \beta }
\Big( \{ \partial_{\alpha} \alpha, \partial_{\beta} b\}-[\partial_\alpha a, \partial_\beta \xi]\Big)\nonumber\\
&+& i [\epsilon^{(1)}, b ] -i\{\xi,A^{(1)}\}+ D \Xi^{(1)}\nonumber\\
 \label{SW82}
\end{eqnarray}

\noindent where $D \epsilon^{(1)} $ and $D \Xi^{(1)}$ represent now   ordinary covariant derivatives as defined in
(\ref{2.4}).
After inserting (\ref{SW6}) into the first of equations (\ref{SW82}) it is possible to find the general solution of the Seiberg-Witten map for the connection  $A$  as \cite{AK}

\begin{eqnarray}
A&=& a-\frac{1}{4}\theta^{\alpha\beta}\left\{a_\alpha,2\partial_\beta a
-D a_\beta\right\}\nonumber\\ 
&+&\theta^{\alpha\beta}D\Big(\sigma f_{\alpha\beta}
+ {\lambda\over 2} [ a_\alpha , a_\beta ]\Big)
+ O(\theta^2)
\nonumber\\
\label{SW91}
\end{eqnarray}

\noindent 
where $\lambda$ appears in (\ref{SW6}) and $\sigma$ is a second parameter associated with
the homogeneous part of the first of equations (\ref{SW82}).
Observe that  the indices associated with the noncommutativity appear explicitly, what be expected
since the Lorentz invariance is broken by the Moyal structure.The covariance associated with the form structure is  however kept. This means that
it is not necessary to write the forms components in   equations  (\ref{SW4}-\ref{SW91})
since they do not mix with the noncommutative structure.
Theses features will also appear  in the following.

It is possible to show  from (\ref{SW6}), (\ref{SW7}) and the second one of equations (\ref{SW82})
 that

\begin{eqnarray}
\Omega^{(1)}[y] &=& -\frac{1}{4} \theta^{\alpha \beta} \{ a_{\alpha },
(\partial_{\beta} +D_\beta) \omega] \}\nonumber\\
&+&\frac{i}{2}\lambda\theta^{\alpha \beta}\big[a_\alpha[a_\beta,\omega]\big]
\label{SW92}
\end{eqnarray}

\noindent is the desired solution for  the  compensating 1-from field. 
To solve the third  equation in (\ref{SW82}) for $B^{(1)}$,  we need to consider the already derived expressions for $\epsilon^{(1)}$,
$\Xi^{(1)}$ and  $A^{(1)}$ given above. As can be inferred from them,
it is not an easy task to achieve a complete solution for $B^{(1)}$ following this route.

However, if we consider the equation defining the gauge variation for $\hat B$ in (\ref{SW82}), we find
a much simpler mapping equation given by

\begin{eqnarray}
\bar\delta\hat B^{(1)}-i[\alpha,\hat B^{(1)} ]&=&-\frac{1}{2}\theta^{\alpha\beta}
\{\partial_\alpha\,\alpha,\partial_\beta \hat b\}
\nonumber\\
&+&i[ \epsilon^{(1)},\hat b]
\label{SW93}
\end{eqnarray}

\noindent whose general solution, when one keeps the form covariance in the sense discussed above,
is given by

\begin{eqnarray}
\hat B^{(1)}&=&-\frac{1}{4}\theta^{\alpha\beta}\left\{a_\alpha,(\partial_\beta +D_\beta) \hat b\right\}\nonumber\\ 
&+&\theta^{\alpha\beta}\Big(\rho[\hat b,f_{\alpha\beta}]
-i{\lambda\over 2} \big[\hat b,[ a_\alpha , a_\beta ]\big]\Big)
\label{SW94}
\end{eqnarray}

\noindent where $\rho$ is a new parameter associated with the homogeneous part of (\ref{SW93}).
Now remembering that
$\hat B=B+D\Omega$ from (\ref{2.241}), one can verify that

\begin{eqnarray}
 B^{(1)}&=&\hat B^{(1)}-D\Omega^{(1)}+i\{ A^{(1)}, \omega\}\nonumber\\
&-&\frac{1}{2}\theta^{\alpha\beta}[\partial_\alpha a, \partial_\beta \omega]
\label{SW95}
\end{eqnarray}

\noindent Inserting (\ref{SW94}) and the expressions for $\Omega^{(1)}$ and $ A^{(1)}$  given in (\ref{SW92})
and (\ref{SW91}) in the above expression,
it is  simple to obtain the complete expression for $B^{(1)}$.

For completeness, we note also that \cite{AK}

\begin{eqnarray}
F^{(1)} &=& 
 \frac{1}{2} \theta^{\alpha \beta}\Big(\{Da_\alpha-\partial_\alpha a,Da_\beta-\partial_\beta a\}\nonumber\\
&-&\frac{1}{2}\{a_\alpha,(\partial_\beta+D_\beta)f\}
+i\sigma[f_{\alpha\beta},f]\nonumber\\
&-&\frac{i}{2}\lambda\big[f,[a_\alpha,a_\beta]\big]\Big)
\label{SW10}
\end{eqnarray}

\noindent and  

\begin{equation}
\hat G^{(1)} =  D\hat B^{(1)} - i [A^{(1)}, \hat b]+ \frac{1}{2}\theta^{\alpha \beta} \{\partial_{\alpha} \hat b,\partial_{\beta}a\} 
\label{SW11}
\end{equation}

Now the action (\ref{2.25}) is mapped to 

\begin{equation}
\begin{array}{lll}
S  = 2\,Tr \int d^4 x & \left( \right. & \frac{1}{2}f  \; \; ^{^{\diamond}}\! \! \! \! \! \! \! \wedge f - \frac{1}{2} \hat g  \; \; ^{^{\diamond}}\! \! \! \! \! \! \! \wedge \hat g + m f \wedge \hat b + 
f  \; \; ^{^{\diamond}}\! \! \! \! \! \! \! \wedge F^{(1)} \\
& & \\
& & \left.  - \hat g  \; \; ^{^{\diamond}}\! \! \! \! \! \! \! \wedge \hat G^{(1)} +
m f \wedge \hat B^{(1)} + m F^{(1)} \wedge \hat b \right)
\end{array}
\label{Action}
\end{equation}

\noindent up to $O(\theta^2)$ terms, 
where $F^{(1)}$, $B^{(1)}$ and $G^{(1)}$ are given above. It can be verified that action (\ref{Action}) is 
invariant under the set of ordinary gauge transformations (\ref{2.14}), since the  Noether identities are kept
under the Seiberg-Witten map.

\section{Conclusions}

We have studied in this work a noncommutative
formulation of the U(N)
Cremmer-\-Scherk-\-Kalb-\-Ramond  theory. The  gauge structure 
has been considered in detail, as well as  a  noncommutative differential form structure
appropriated to describe the model. It was also constructed the Seiberg-Witten map between 
this noncommutative gauge theory
and its ordinary counterpart, in first order in the 
noncommutative parameter $\theta$. In this last subject,
not only the vector gauge sector has been considered, but also the rank two gauge algebra
associated with the antisymmetric tensor gauge fields. We observe that
it has not been  explored here some characteristic features usually associated with the ordinary CSKR
model such as the effective topological mass generation for the vectorial sector, since they  constitute direct 
generalizations of the ordinary case. Specific features associated with noncommutative quantum field theories \cite{REVIEW}
such as renormalizability, broken of unitarity, presence of anomalies or mixing of ultraviolet and infrared divergencies have been left for consideration in a future work.

\vspace{1 true cm}

{\bf Acknowledgments:} This work is supported in part by CAPES and CNPq (Brazilian research agencies).

\end{document}